\documentclass[utf8]{FrontiersinPreprint} 

\usepackage{url,hyperref,lineno,microtype,subcaption}
\usepackage[onehalfspacing]{setspace}

\linenumbers

\usepackage{xcolor}
\usepackage{siunitx}
\usepackage{circuitikz}
\usepackage{mathtools} 
\usepackage{suffix}  
\usepackage{xstring} 
\usepackage[nohyperlinks, printonlyused]{acronym}

\newcommand{\Z}[1]{\underline{Z}_{\mathrm{#1}}}
\newcommand{\Zk}[1]{\underline{Z}_{#1}}

\usepackage{tikz}
\usepackage{pgfplots}
\usetikzlibrary{calc,arrows.meta,positioning,fillbetween}
\tikzset{pmd/.style={%
\circuitikzbasekey/.cd,%
resistors/scale=0.7,%
capacitors/scale=0.7,%
},
}
\pgfplotsset{
    compat=1.18,
    GenPlotStyle/.style={
    width=\linewidth,
    height=0.9\linewidth,
    ymajorgrids=true,
    grid style=dashed,
    every axis plot/.append style={very thick}
    },
}

\def\keyFont{\fontsize{8}{11}\helveticabold }
\def\firstAuthorLast{Puchtler {et~al.}} 
\def\Authors{Steffen Puchtler\,$^{1,*}$,Julius van der Kuip\,$^{1}$,Florian Michael Becker-Dombrowsky\,$^{1}$ and Eckhard Kirchner\,$^{1}$}

\def\Address{$^{1}$Technical University of Darmstadt, Institute for Product Development and Machine Elements, Darmstadt, Germany}
\def\corrAuthor{Otto-Berndt-Str. 2, 64287 Darmstadt, Germany}

\def\corrEmail{office@pmd.tu-darmstadt.de}

\begin{document}
\nolinenumbers

\onecolumn
\firstpage{1}

\title[Impedance Measurement of Rolling Bearings]{Impedance Measurement of Rolling Bearings Using an unbalanced AC Wheatstone Bridge} 

\author[\firstAuthorLast ]{\Authors} 
\address{} 
\correspondance{} 

\extraAuth{}

\maketitle

\begin{abstract}

\section{}
Industry 4.0 drives the demand for cost-efficient and reliable process data and condition monitoring. Therefore, visualizing the state of tribological contacts becomes important, as they are regularly found in the center of many applications. Utilizing rolling element bearings as sensors and monitoring their health by the electrical impedance method are promising approaches as it allows e.g. load sensing and detection of bearing failures. The impedance cannot be measured directly, but there are various methods available. This work discusses advantages and disadvantages and suggests the AC Wheatstone bridge as a reliable way of measuring impedances with low phase angles at sampling rates in the \unit{\kilo\hertz} range. The corresponding equations are introduced, a simulation built, an uncertainty mode and effects analysis carried out and sample measurement results of real rolling elements shown. It can be demonstrated that the AC Wheatstone bridge meets the proposed requirements for sensory utilization and condition monitoring when the bearing is operated in the hydrodynamic regime.

\tiny
 \keyFont{ \section{Keywords:} bearing, impedance, measurement, visualization, condition monitoring} 
\end{abstract}

\section{Introduction}

\subsection{Impedance measurement at rolling bearings}
Rolling bearings are central components in many machines. Because of their crucial role in leading process forces into the machine structure, bearing failures cause nearly \SI{20}{\percent} of all machine breakdowns \citep{Dahlke.1994,SchaefflerTechnologiesAG&Co.KG.2000}. With the arising e-mobility, another bearing fault type gets into the focus of research, namely surface damages based on harmful electric bearing currents \citep{Harder.2022b}. In both cases, the electric impedance can give information about the bearing condition. 
\newline To avoid unexpected system faults of machines, a precise knowledge about the mechanical and electrical bearing loads as well as health conditions is essential. Martin et al. could show that measuring the bearing impedance is able to detect damages in the runway surfaces like pittings \citep{Martin.2022}. Becker-Dombrowsky et al. used the data generated by Martin et al. and investigated them in time and frequency domain. Three different phases in rolling bearing lifetime were identified and the end of bearings' life could be detected \citep{BeckerDombrowsky.2023}. So, impedance based condition monitoring of rolling element bearings is possible. The gauging principle is based on the description of the electrical behavior of the rolling contact, which is regarded in the next section. 

\subsection{Electrical behavior of rolling contact}
Depending on the lubrication film thickness, three different electrical equivalent models can be described. In case of dry friction, a direct metallic contact between rolling elements and runways exists. The observed electrical behavior is equivalent to an ohmic resistance. With increasing lubrication film thickness, an \ac{ehl} contact is built between the rolling elements and the runways. Now, the equivalent circuit can be extended by a capacitor in parallel connection. The capacitor is the Hertz'ian contact zone, which is formed by the bearing loads \citep{Prashad.1988, Muetze.28.10.2003, Gemeinder.2016, Schneider.2022, Schneider.2022b}. If the lubrication film thickness is insufficient, breakdown voltages like in EDM machines harm the surfaces of the contact partners. The reason is, that the bearing current density in the Hertz'ian area is high enough to overbear the lubricant resistance in the \ac{ehl} contact, so a harmful current can be transmitted through the lubrication gap. When a sufficient lubrication film separates the contact partners completely, the \ac{ehl} contact zone can be modeled as the described parallel connection of ohmic resistance and capacitor, in which the capacitive part prevails \citep{Prashad.1988, Muetze.28.10.2003, Gemeinder.2016, Schneider.2022, Schneider.2022b}. The rolling bearing impedance is a result of the impedance of the single rolling element contact to the runway. These single contacts are connected in parallel. The single contact itself is modeled as a serial connection of two plate capacitors, each in parallel connection to an ohmic resistance. Knowing the electrical properties of rolling element bearings enables their usage as sensors or their usage as condition monitoring device. To do so, the impedance has to be measured during the operation of the bearing. In the next section, impedance measurement approaches will be presented.

\subsection{Impedance measurement methods}
Currently, impedance measurement is used for lubrication film thickness observation \citep{Barz.1996}. Martin et al. could show the possibilities of using the electric bearing properties for condition monitoring. They were able to observe the pitting growth after initial damage in the rolling bearing raceways and to gauge the damage dimensions \citep{Martin.2022}. Becker-Dombrowsky et al. expanded the analysis of Martin et al. in the time and frequency domain to enlarge the amount of data using feature engineering techniques \citep{BeckerDombrowsky.2023}. They are able to distinguish three rolling bearing operational time phases and to detect bearings end of life \citep{BeckerDombrowsky.2023}. The impedance measurement can also be used to identify the load condition in the bearing \citep{Schirra.2018}. So, the rolling bearing becomes an active element in the measurement circle. When the impedance is used to explain harmful electric conditions in the rolling contact, it is just a reacting or passive element. For both approaches, the Hertz'ian contact area has to be calculated as follows: 

\begin{equation}
C_\mathrm{Hertz} = \varepsilon_0 \cdot \varepsilon_\mathrm{r} \cdot \frac{A_\mathrm{Hertz}}{h_\mathrm{c}},
\label{eq:Capa1}
\end{equation}

where $C_\mathrm{Hertz}$ is the measured capacitance, $\varepsilon_0$ the vacuum permittivity, $\varepsilon_r$ the relative permittivity, $A_\mathrm{Hertz}$ the Hertz'ian contact area and $h_{c}$ the lubrication film thickness in the \ac{ehl} contact. Approaches for the calculation of ther capacitance contributors are discussed by \cite{Puchtler.2022}. Based on the Hertz'ian theory, the load in the rolling contact can be calculated, which enables more suitable durability analysis and failure monitoring \citep{Schirra.2019}. Maruyama et al. are using the impedance measurement approach to detect lubricants condition in rolling contact \citep{Maruyama.2023}. This application field is not addressed in this article. The following section gives a short introduction into different impedance measurement approaches. 

\subsubsection{Charging Curve Detection}
This measurement method is used for lubrication film thickness detection. It utilizes direct current as a carrier signal. To get the impedance of the \ac{ehl} contact, the known current is let through the bearing. Then the time is measured until a defined voltage level is reached. With this information, the capacitance and the impedance can be calculated. The capacitance is proportional to the time \citep{Barz.1996}. In case of metallic contacts, breakdown currents occur, which influence the measurement results. In the experiments of \cite{Wittek.2015}, beside the metallic contacts, exceeding the breakdown voltage in the contact zone caused breakdown currents. These currents can be harmful to bearings, so the charging curve detection is not suitable for a condition monitoring system.

\subsubsection{Current and voltage measurement}
This gauging method is based on the Ohm's law for the \ac{ac} circuit. By measuring the bearing voltage over the time and the current through the bearing, the impedance can be calculated \citep{KeysightTechnologies.2020}. It is applicable for measurement frequencies between \SI{10}{\kilo\hertz} and \SI{100}{\mega\hertz}. Its accuracy depends on the voltage and current measurement systems, which is defined by the working frequency of the gauging device \citep{KeysightTechnologies.2020}. The measurement errors can be corrected or minimized, when the inner ohmic resistance of the device is lower than the measured impedance. So, this method is mainly used for detecting higher impedances \citep{Plamann.2016}. 

\subsubsection{Voltage comparison}
The voltage comparison is able to minimize measurement errors, because it is based on a voltage divider. That means, it needs a known reference impedance $\Z{ref}$ to detect the unknown impedance $\Z{x}$. The voltages over the reference impedance $\underline{v}_\mathrm{ref}$ and the generator voltage $v_\mathrm{gen}$ are measured and the searched impedance is calculated by following equation: 
\begin{equation}
\Z{x} = \left(\frac{v_\mathrm{gen}}{\underline{v}_\mathrm{ref}}-1\right) \Z{ref}.
\label{eq:MeasMethEqu1}
\end{equation}
The equivalent circuit is shown in Figure~\ref{fig:voltageComparison_circuit}. The advantage of this method is, that systematic errors will be minimized \citep{Lerch.2016}. But the inner resistance of the voltage gauging device has to be higher than the observed impedance. If it is not possible to realize this, the reference impedance has to be in the same range as the observed impedance to reduce measurement errors.

\subsubsection{Measurement Bridge}\label{sec:1MeasurementBridge}
Measurement bridges are able to use direct and alternating current as carrier signal. They need three known reference impedances to detect the unknown fourth impedance. The generator voltage and the bridge voltage are measured using an oscilloscope or equivalent device \citep{Lerch.2016}. The unknown impedance is calculated by transforming the equation written below for the searched impedance \citep{Lerch.2016}: 
\begin{equation}
\underline{v}_\mathrm{m} = \frac{\Z{x}\Z{2}-\Z{1}\Z{3}}{(\Z{1}+\Z{x})(\Z{2}+\Z{3})}v_\mathrm{gen}.
\label{eq:MeasMethBridge1}
\end{equation}
An equivalent circuit of a measurement bridge is shown in Figure~\ref{fig:ac-bridge_circuit}. The inner resistance of the measurement device and the resistance of the wires can lead to errors, which has to be considered. The main advantage of this measurement principle is its robustness against environmental influences \citep{Plamann.2016} and its sensitivity \citep{Mitvalsky.1964}. According to the measurement task, an AC Wheatstone bridge can be configured in different ways \citep{Cone.1920}. Another variant is the balanced AC Wheatstone bridge, where a reference impedance is tuned until the bridge voltage is zero $\underline{v}_\mathrm{m} = \SI{0}{\volt}$ \citep{Takagishi.1980}. For further research into AC Wheatstone method, \cite{Gupta.2020} and \cite{Rybski.2015} and for its application, \cite{Chattopadhyay.2012} should be mentioned.

\begin{subfigure}
    \setcounter{subfigure}{0}
    \centering
    \begin{minipage}[t]{0.49\textwidth}
        \begin{circuitikz}[pmd, european resistors]
        \draw
          (0,0) to [short, -] (2,0)
          (0,4) to [short, -] (2,4)
          (0,0) to [vsourcesin, l=$v_\mathrm{gen}$] (0,4)
          (0,0) node[ground](GND){}

          (2,0) to [R, l=$\Z{x}$] (2,2)
          to [R, l=$\Z{ref}$, -] (2,4)

          (2,2) to [short, -o] (3.5,2)
          (2,4) to [short, -o] (3.5,4)
          (3.7,4) to [open, v^=$\underline{v}_\mathrm{ref}$] (3.7,2)
          ;
        \draw [color = black!60]
          (3.5,2) to [R, l=$\Z{m}$] (3.5,4)
        ;
        \draw(-0.5,4) node[anchor=  0]{\textbf{(A)}};
        \end{circuitikz}
        \caption{Circuit diagram of the voltage comparison method.}
        \label{fig:voltageComparison_circuit}
    \end{minipage}  
    \hfill
\setcounter{subfigure}{1}
    \begin{minipage}[t]{0.49\textwidth}
        \begin{circuitikz}[pmd, european resistors]
        \draw
          (0,0) to [short, -] (5,0)
          (0,4) to [short, -] (5,4)
          (0,0) to [vsourcesin, l=$v_\mathrm{gen}$] (0,4)
          to [short, -o] ++(-0.5,0)
          (0,0) node[ground](GND){}
          to [short, -o] ++(-0.5,0)
    
          (2,0) to [R, l=$Z_\mathrm{x}$] (2,2)
          to [short, -o] ++(0.5,0)
          
          (2,2) to [R, l=$\Z{1}$] (2,4)
          (5,2) to [R, l=$\Z{2}$] (5,4)
          (5,0) to [R, l=$\Z{3}$] (5,2)
          to [short, -o] ++(-0.5,0)
          (2.5,1.7) to [open, v>=$\underline{v}_\mathrm{m}$] (4.5,1.7)
        ;
        \draw [color = black!60]
          (2.5,2) to [R, l=$\Z{m}$] (4.5,2)
        ;
        \draw [color = blue]
          (-0.5,0) node[anchor=  0]{0}
          (-0.5,4) node[anchor=  0]{1}
          ( 2.5,2) node[anchor=-90]{2}
          ( 4.5,2) node[anchor=-90]{3}
        ;
        \draw(-1,4) node[anchor=  0]{\textbf{(B)}};
        \end{circuitikz}  
        \caption{Circuit diagram of the AC Wheatstone bridge.}
        \label{fig:ac-bridge_circuit}
    \end{minipage}

\setcounter{subfigure}{-1}
    \caption{Circuit diagrams of  \textbf{(A)} the voltage comparison method and \textbf{(B)} the AC Wheatstone bridge.}
    \label{fig:circuitDiagrams}
\end{subfigure}

\subsubsection{Comparison of measurement methods}
In this subsection, the suitability of the single impedance measurement methods for the single contact measurement and the rolling bearing operational time observation will be discussed and two approaches chosen. 

The charging curve detection can lead to harmful bearing currents over the observation time, so it is possible that the bearing will be damaged by this measurement method. Because of the discharge currents when metallic contacts occur, a constant signal during the operational time will not be possible. So, the charging curve detection is not suitable for the aim of this paper. 

The current and voltage measurement method needs a lower inner ohmic resistance of the measurement devices than the observed impedance to reduce measurement errors. The inner resistance of the used oscilloscope is higher than one mega ohm. The rolling bearing impedance for a carrier signal of \SI{20}{\kilo\hertz} is about several kilo ohm, which means it is lower than the inner resistance. So this method is not applicable for this case. 

The voltage comparison has already been used in early research to detect rolling bearing damages \citep{Martin.2022}. It used the bearing seat isolation of the test rig chamber as a reference impedance, which can be influenced by the environment and the tests themselves. The carrier signal was about \SI{2.5}{\mega\hertz}. This reference impedance can not be adapted to the expected rolling bearing impedance, so a possible measurement error can not be reduced. An external reference impedance could be adapted and avoid measurement errors. This would be able with a measurement bridge. Because of its robustness against environmental influences, an alternating voltage measurement bridge is more suitable for the single contact and operational time observation. So the measurement bridge is chosen as a new impedance measurement approach. It will be compared to the voltage comparison method in this paper as state of the art in the next sections.

\section{Materials and Methods}

\subsection{Alternating Voltage Measurement Bridge Function}
\label{sec:measurment_bridge_function}

In the following, the AC Wheatstone bridge as described in Section~\ref{sec:1MeasurementBridge} and Figure~\ref{fig:ac-bridge_circuit} is enhanced. To take the loading effect of the differential oscilloscope probe measuring $\underline{v}_\mathrm{m}$ into account, the probe's impedance $\Z{m}$ is introduced. Then, solving for $\Z{m}$ yields

\begin{equation}\label{eq:ac_bridgeZ_x}
	\Z{x} = \Z{1} \frac{\Z{3}\Z{m}+[(\Z{2}+\Z{3}) \Z{m}+\Z{2}\Z{3}]\underline{v}_\mathrm{m}/v_\mathrm{gen}}
	                             {\Z{2}\Z{m} -[(\Z{2}+\Z{3})(\Z{m}+\Z{1}) +1 ]\underline{v}_\mathrm{m}/v_\mathrm{gen}} .
\end{equation}

The generator voltage phase is defined as $\operatorname{arg}(v_\mathrm{gen}) = 0^\circ$ and therefore, $v_\mathrm{gen}$ is a real number. $\Z{1}$, $\Z{2}$, $\Z{3}$ and $\Z{m}$ are known and considered constant. Thus, only the complex voltage ratio $\underline{v}_\mathrm{m}/v_\mathrm{gen}$ must be measured during the experiment. This is done by a digital differential oscilloscope with the sampling rate $f_\mathrm{m}$ or sampling interval $T_\mathrm{m}=1/f_\mathrm{m}$. A subscript $k$ will be assigned to each sample. A  phase information cannot be derived reliably from a single sample, therefore, a moving sum is used which covers a number of $N$ samples symmetrically around $k$. For the calculation of the imaginary part of $\underline{v}_\mathrm{m}/v_\mathrm{gen}$, a $90^\circ$ phase shifted signal of the input voltage $v^*_\mathrm{gen}$ is calculated.

\begin{gather}
	\mathrm{Re}\{\underline{v}_{\mathrm{m},k}/v_{\mathrm{gen},k}\} = \frac{2}{N \hat{v}^2} \sum_{i=k-N/2}^{k+N/2} \underline{v}_{\mathrm{m},i} v_{\mathrm{gen},i},\\
	\mathrm{Im}\{\underline{v}_{\mathrm{m},k}/v_{\mathrm{gen},k}\} = \frac{2}{N \hat{v}^2} \sum_{i=k-N/2}^{k+N/2} \underline{v}_{\mathrm{m},i} v^*_{\mathrm{gen},i}.
\end{gather}

Subsequently, a first order low pass filter at $f_\mathrm{m}/2$ is applied to remove any remaining ripple and to avoid any false interpretation of the signal, as higher frequencies will not be reliably measured. To reduce the effect of any stray capacitances and cable inductances, an open-short adjustment is carried out. Therefore, the impedance is measured with the \ac{dut} removed ($\Z{open}$) and shorted ($\Z{short}$). These values are used to retrieve the \ac{dut}'s impedance from the measured impedance $\Z{x}$, Equation~(\ref{eq:ac_bridgeZ_x}), \citep{Muhl.2020},

\begin{equation}
    \Z{DUT} = \frac{\Z{x}-\Z{short}}{1-(\Z{x}-\Z{short})/(\Z{open}-\Z{short})}.
\end{equation}

Finally, the capacitance $C_\mathrm{DUT}$ and resistance $R_\mathrm{DUT}$ of the equivalent circuit at the carrier frequency $\omega = 2\pi f_\mathrm{gen}$ can be calculated as

\begin{gather}
    C_\mathrm{DUT} = -\frac{\mathrm{Im}\{\Z{DUT}\}}{\omega \, |\Z{DUT}|^2},\label{eq:CfromZ}\\
    R_\mathrm{DUT} = \mathrm{Re}\{\Z{DUT}\} + \frac{\mathrm{Im}\{\Z{DUT}\}^2}{\mathrm{Re}\{\Z{DUT}\}}.\label{eq:RfromZ}
\end{gather}

\subsubsection{Measurement Uncertainty}\label{sec:measurementUnsertainty}
The impedances $\Z{1}$, $\Z{2}$, $\Z{3}$ and $\Z{m}$ cannot be measured directly once the bridge is assembled and connected to the oscilloscope. As these cables contribute to the impedances, they cannot be measured before assembly. The assembled bridge is a four-terminal circuit, hence, the six impedances between the terminals can be calculated from six measurements between different terminals each. The four poles are denominated as $0$, $1$, $2$ and $3$, cf. Figure~\ref{fig:ac-bridge_circuit}. The impedance $\Zk{ij}$ describes the single impedance between poles $i$ and $j$. The impedance $\Zk{\mathrm{m},ij}$ describes the measured capacitance between the poles $i$ and $j$ including the effects of all other impedances in the network,

\begin{equation}
\Zk{\mathrm{m},ij} = \frac{\Zk{ij} \left[ \Zk{jk} \Zk{jl} (\Zk{ik} + \Zk{kl} + \Zk{il}) + \Zk{ik} \Zk{kl} (\Zk{jl} + \Zk{jk} + \Zk{il}) \right]}{\splitfrac{\Zk{ij} [(\Zk{jk}+\Zk{jl}) (\Zk{ik} + \Zk{kl} + \Zk{il})] + \Zk{jk} \Zk{jl} (\Zk{ik} + \Zk{kl} }{+ \Zk{il}) + \Zk{ik} \Zk{kl} (\Zk{ij} + \Zk{jk} + \Zk{jl} + \Zk{il})}} \; \mathrm{with} \; i,j,k,l \in \{0,1,2,3\}.
\end{equation}

Using the trust-region dogleg algorithm implemented as fsolve in Matlab, all $\Zk{ij}$ can be retrieved by the measured $\Zk{\mathrm{m},ij}$. For the AC Wheatstone bridge used, the following values were retrieved, corresponding to an RC-equivalent according to Equations~(\ref{eq:CfromZ}) and (\ref{eq:RfromZ}), 
\begin{itemize}
    \item $\Z{12} \quad \Rightarrow$ \quad
          $C_1         =\SI{  990.56(3)}{\pico\farad}$ \qquad 
          $R_1         =\SI{ 15805(879)}{\kilo\ohm}$
    \item $\Z{13} \quad \Rightarrow$ \quad
          $C_2         =\SI{  989.29(6)}{\pico\farad}$ \qquad 
          $R_2         =\SI{11594(1034)}{\kilo\ohm}$
    \item $\Z{03} \quad \Rightarrow$ \quad
          $C_3         =\SI{ 1059.54(2)}{\pico\farad}$ \qquad 
          $R_3         =\SI{  890   (2)}{\kilo\ohm}$
    \item $\Z{23} \quad \Rightarrow$ \quad
          $C_\mathrm{m}=\SI{    4.22(4)}{\pico\farad}$ \qquad 
          $R_\mathrm{m}=\SI{28236(5700)}{\kilo\ohm}$
\end{itemize}

The impedance $\Z{01}$ is parallel to the power source and therefore not considered. $\Z{02}$ is parallel to the \ac{dut} and will be therefore considered by the open measurement. In case, the open measurement cannot be carried out, this value can be used as $\Z{open}$.

\subsection{Uncertainty Consideration}\label{sec:uncertaintyConsideration}
\begin{figure}[h!]
\begin{center}
\includegraphics[width=\linewidth]{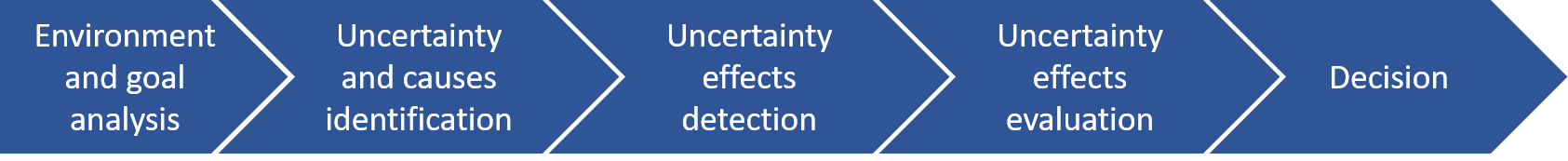}
\end{center}
\caption{\ac{umea} approach process.}\label{fig:UMEA}
\end{figure}

The \ac{umea} methodology, is used for the analysis of uncertainties in the context of product development. It is divided into five steps, which are briefly described below. Figure~\ref{fig:UMEA} schematically shows the procedure for this methodology.

The first step is an environment and target analysis, in which the systems to be considered are defined by means of system boundaries and differentiated from other systems \citep{Engelhardt.2012}. Interfaces to other systems are named and their influences on the system under consideration are taken into account. 
This is followed by the identification of uncertainties, in which these are named, and their causes are investigated. For this, system models are necessary in order to be able to describe the influences sufficiently precisely \citep{Engelhardt.2012}. 
Then the uncertainty effects are determined, which is followed by the penultimate step of the \ac{umea} methodology, the uncertainty evaluation. Based on this, a final decision is made regarding the countermeasures to be applied \citep{Engelhardt.2012}. 

First, the investigated system is described. On the basis of this description, the necessary system boundaries are defined and the disturbance, input, output and secondary variables acting on it are identified. The definitions of these quantities given below follow that of Mathias and are briefly presented here \citep{Mathias.2016}.

\begin{itemize}
    \item 
    The \textbf{input variable} is a desired quantity which is converted into an output quantity based on the system behavior. Thereby real input quantities can differ from the desired ones. 
    \item
    An \textbf{output variable} is a desired variable that is formed from the input variables and the disturbance variables as a result of the system behavior. The goal is that the system produces an output variable that corresponds to the desired output variable. Deviations are possible in this regard.
    \item
    \textbf{Disturbance} variables represent undesirable effects on the system, which can thus cause deviations in the output variables. 
    \item
    \textbf{Secondary variables} are unwanted variables that result from the system behavior. They can in turn act as disturbance variables on neighboring systems. 
\end{itemize}

As can be seen from the definitions, disturbance variables and secondary variables have largely the same effect and can therefore be regarded as equivalent \citep{Mathias.2016}. Thus, no distinction is made between them in the identification of disturbances by means of lists.

\subsection{Simulation Model}\label{sec:simulationModel}
\begin{figure}[h!]
\begin{center}
\includegraphics[width=\linewidth]{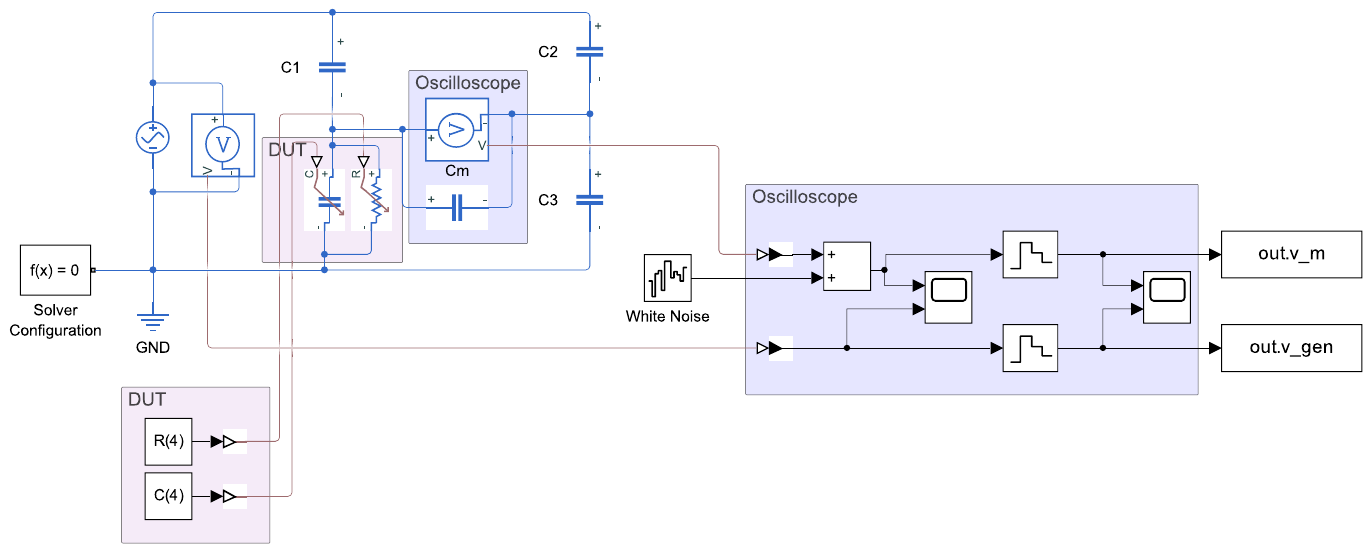}
\end{center}
\caption{Matlab Simulink simulation model of the AC Wheatstone bridge.}\label{fig:simModel}
\end{figure}

A Matlab Simulink model was developed to investigate the behavior of the AC Wheatstone brige in ideal conditions. Figure~\ref{fig:simModel} shows the diagram of the model. The parallel connection of a variable capacitor and a resistor simulates the \ac{dut}. The impedances $\Z{1}$, $\Z{2}$ and $\Z{3}$ are modeled using a real capacitor with a parallel resistance of the values retrieved in Section~\ref{sec:measurementUnsertainty}. Similarly, the oscilloscope is modeled by an ideal voltage sensor in parallel with a real capacitor. The voltage signals of bridge voltage $v_\mathrm{m}$ and the generator $v_\mathrm{gen}$ are then sampled with the sampling rate $f_\mathrm{s}$ and resolution (\SI{12}{\bit}) of the oscilloscope used. For some tests, a white noise is added to the measured bridge voltage simulating electric interference. The resulting signal is then processed with the same scripts later used for measurements in the real application. The simulation of \SI{10}{\milli\second} is then carried out using Matlab's ode14x solver with a fixed step size of $0.05/f_\mathrm{s}$.

\subsection{Application Tests}
\subsubsection{Rolling Bearing Test Rig}\label{sec:testRig}
\begin{figure}[h!]
\begin{center}
\includegraphics[width=.7\linewidth]{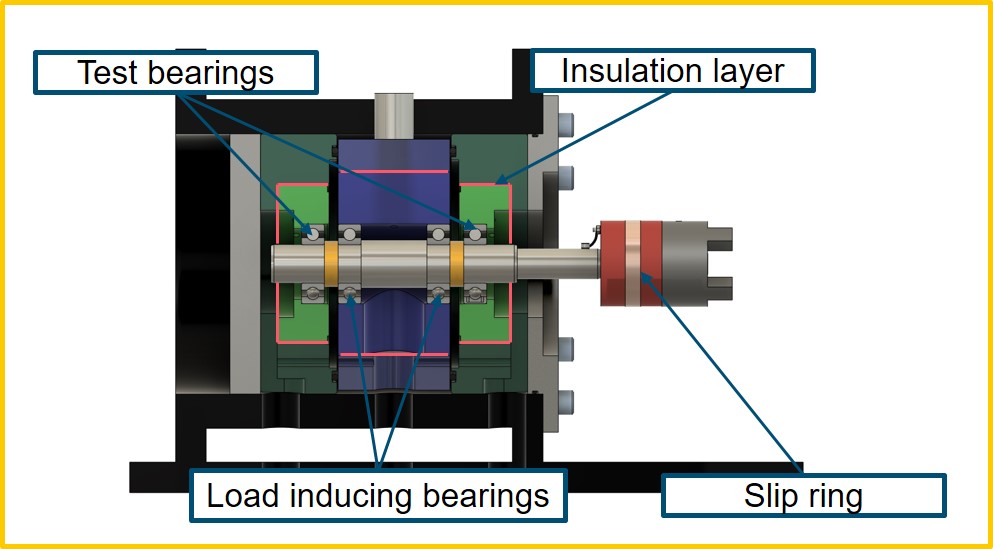}
\end{center}
\caption{Cross section of one rolling bearing test rig chamber.}\label{fig:CStestChamber}
\end{figure}

The rolling bearing test rig has four test chambers which can be operated independently. Four rolling element bearings are run in a single chamber. All bearings are loaded radially and in addition, two bearings can be loaded axially if necessary. Each chamber is equipped with an own recirculating oil lubrication system, so oil lubrication besides grease lubrication can be investigated. For test observation, axial and radial vibration sensors and four temperature sensors are located at the chamber. In addition, the motor speed and torque is monitored to enlarge the system information spectrum. Figure~\ref{fig:CStestChamber} shows a cross-section of one chamber. The two rolling bearings in the center induce the radial loads to the two outer bearings. The outer bearings can be loaded axially by the hydraulic actuator on the left side. The radial load induction is also realized by a hydraulic actuator above the chamber. 
\newline
To avoid parasitic currents and other disturbing factors based on electric phenomena, the steel bearing seats are surrounded by a \SI{50}{\milli\meter} thick isolation layer made from fiber glass. For the electrically investigated bearings, the isolation is over-bridged using contact pins. The shaft is contacted by a slip ring.

\subsubsection{Single Contact Test}
An application of the impedance measurement method described is testing single steel ball bearings. Therefore, a ceramic ball of a hybrid bearing is replaced by a steel ball of the same diameter, as first described by \cite{Jabonka.2018}. Thus, the impedance of a single ball can be measured. Some test results as well as the calculation of the changed mechanical behavior due to different elasticities and thermal expansion coefficients of steel and ceramic, is described in \cite{Puchtler.2023}. For small load angles $\beta=\tan\left(F_\mathrm{a}/F_\mathrm{r}\right)$ a high change of capacitance during an orbit of the steel ball is expected due to the distinct load zone. Thus, a capacitance maximum will appear with each cage revolution. 

\subsubsection{Fatigue Tests}
To investigate the applicability of the new measurement approach, fatigue tests were run at the test rig described in Section~\ref{sec:testRig}. Two different rolling bearing types are used, deep groove ball bearings (6205) and angular deep groove ball bearings (7205). Both bearing types have the same dimensions. All bearings are loaded by a resulting load ratio of $C/P = 1.6$ with changing load angle $\beta$.The rotational speed is stationary at \SI{5000}{rpm}. Fully lubrication is applied using the recirculating oil lubrication system of the test chamber with an oil flow of \SI{10}{\litre \per\minute}.  
The tests are run until the sensor system of the test rig detects a damage. Tests with a duration more than \SI{300}{\hour} are ended and will not be investigated in this research. The results of the fatigue tests are compared to the data of Martin et al. The data quality is important as well as the accuracy based on the calculations. 

\section{Results and Discussion}

\subsection{\ac{umea}}
\label{sec:umea_results}
In this section, the predominant disturbance variables are identified. For this purpose, system boundaries are defined first and then the disturbance variables acting on the respective system are determined. These are evaluated in accordance with the methodology described in Section~\ref{sec:uncertaintyConsideration}. This is followed by an explanation of possible solutions for dealing with the disturbance variables.

\subsubsection{Disturbance identification}
The analyzed system consists of the measuring system for determining the impedance of the test bearings, which includes the signal lines and the oscilloscope. It also includes the connection contacts of the bearing seats and the associated measuring leads. In addition, the lines of the carrier signal, which is generated by a signal generator, and the signal generator itself are counted as part of the system. The measurement system mentioned here must be distinguished from the systems installed in the test bench for test bench operation.

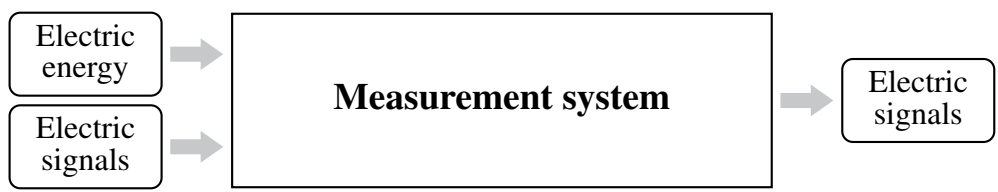
\begin{figure}[h!]
    \centering
    \begin{tikzpicture}[
        node distance= 9mm and 3mm,
        leftIns/.style = {minimum width=20mm,
                        draw, thick, rounded corners, 
                        minimum height=11mm, align=center},
        main/.style = {minimum width=#1,
                        draw, thick, minimum height=23mm,
                        font=\Large\bfseries},
        Arrow/.style = {line width=2mm, draw=gray, 
                        -{Triangle[length=3mm,width=4mm]},
                        shorten >=1mm, shorten <=1mm},
                            ]
        \node[main=71mm] (c) {Measurement system};
        \node[leftIns, left=9mm of c.north west, yshift=-5.5mm] (l1){Electric\\energy};
        \node[leftIns,below=1mm of l1](l2){Electric\\signals};
        \node[leftIns, right=9mm of c.east] (r) {Electric\\signals};
        \draw[Arrow]    (l1) -- (c.west  |- l1);
        \draw[Arrow]    (l2) -- (c.west  |- l2);
        \draw[Arrow]    (c.east  |- r) -- (r);
    \end{tikzpicture}
\caption{Input and output parameters measurement system.}\label{fig:InputOutput}
\end{figure}

The measuring system for impedance measurement considered in this work uses an AC voltage as a carrier signal, from whose changes conclusions are drawn about the states of the test bearings. Thus, this electrical signal represents an input variable for the measuring system. In order to operate the measuring system, components such as the oscilloscope require electrical energy, which therefore also represents an input variable. The measuring system perceives the changes in the impedance, whereby it generates electrical signals as an output variable, which are further processed by software after analog-to-digital conversion. The identified disturbance factors are displayed in Figure~\ref{fig:InputOutput}. 

The determination of external disturbances is based on the use of checklists. The identified disturbance variables are now presented here and their effects will be evaluated.

\subparagraph*{Material-bound waves}
Neighboring test stands and test setups as well as test chambers can result in the transmission of sound waves. These occur in the form of structure-borne sound when they are transmitted via the foundation to the test stand and thus the test station. Furthermore, structure-borne sound can be transmitted to the test chamber through the base plate on which the motor and test chamber are mounted. In addition, structure-borne noise is transmitted between the test stations by the test stand frame. Another possibility is transmission as airborne sound directly to the test station and test chamber. The sound can cause vibrations that can be reflected in the measurement signal. 

\subparagraph*{Electricity and magnetism}
In the systems of the test station, the test chamber and the measuring system, electrical operating currents are present during normal operation, as are current changes and fluctuating electrical voltages. Since all measurement methods in question use a carrier signal in the form of an electrical voltage, this can be affected by changes in current, normal operating currents and voltage changes. 
A greater interference potential results from unequal charge levels. This can result in unintended current flows that affect the measurements as such. Grounding problems are also assigned to this area. 
Galvanic coupling of the signal generator used and the test stand may occur due to a ground loop. This occurs when the respective housings are grounded via the protective contacts of their mains connections because of the protection against contact. As a result, interference currents are induced, which can directly influence the carrier signal and therefore also the measurement. 
Since electric motors are used to generate the speeds required for the tests, electromagnetic fields can also occur which affect the measuring system. Likewise, influences by neighboring test stands, which work with stronger electromagnetic fields, are possible. In order to be able to make more detailed statements on this, field measurements must be carried out in the test hall.

\subparagraph*{Motion-based phenomena}
Because the test rig is located in the Earth's gravitational field, Newtonian gravity is present for all systems. However, this has no direct effect on the measurements and affects all systems equally. Thus, gravity is not considered further as a disturbance variable.

\subparagraph*{Foreign solids and liquids}
Due to dusts in the ambient air, deposits of these may occur on the system under consideration. The same applies to greases and oils. Since the measuring system is not encapsulated from the environment, it is subject to the humidity of the environment. 
All the factors mentioned have an equal effect on the system, and their effect on the measurement must be questioned. Nevertheless, there is a risk of measurement system failure due to ingress of foreign matter such as dust particles into the measurement equipment. However, these problems are usually solved on the manufacturer's side, so they will not be discussed in detail here. 
In the case of slip ring contacting, contamination up to the formation of an oxide layer can occur. These can influence the measurement signal. 

\subparagraph*{Thermodynamics}
Due to the environment and the operation of the test stand itself, the phenomena of heat dissipation and heat supply occur. These can result in changes in resistance, which can lead to falsification of measurement results. 
The following Figure~\ref{fig:DisturbanceFactors} summarizes the findings with regard to the disturbance variables acting on the measurement system.

\begin{figure}[h!]
    \centering
    \begin{tikzpicture}[
        node distance= 9mm and 3mm,
        upIns/.style = {minimum width=35mm,
                        draw, thick, rounded corners, 
                        minimum height=11mm, align=center, rotate=90},
        leftIns/.style = {minimum width=20mm,
                        draw, thick, rounded corners, 
                        minimum height=11mm, align=center},
        main/.style = {minimum width=#1,
                        draw, thick, minimum height=23mm,
                        font=\Large\bfseries},
        Arrow/.style = {line width=2mm, draw=gray, 
                        -{Triangle[length=3mm,width=4mm]},
                        shorten >=1mm, shorten <=1mm},
                            ]
        \node[upIns]             (a) {Structure-borne\\ \& airborne sound};
        \node[upIns,right=6mm of a.south west]  (a2){Changes of\\operating current};
        \node[upIns,right=6mm of a2.south west] (a3){Galvanic couple};
        \node[upIns,right=6mm of a3.south west] (a4){Electromagnetic\\fields};
        \node[upIns,right=6mm of a4.south west] (a5){Pollution};
        \node[upIns,right=6mm of a5.south west] (b) {Temperature\\changes};
        \path   let \p1 = (a.north),
                    \p2 = (b.south),
                    \n1 = {veclen(\x2-\x1,\y2-\y1)} in 
                node[main=\n1,
                     below right=9mm and 0mm of a.north west] (c) {Measurement system};
        \node[leftIns, left=9mm of c.north west, yshift=-5.5mm] (l1){Electric\\energy};
        \node[leftIns,below=1mm of l1](l2){Electric\\signals};
        \node[leftIns, right=9mm of c.east] (r) {Electric\\signals};
        \draw[Arrow]    (a)  -- (a  |- c.north);
        \draw[Arrow]    (a2) -- (a2 |- c.north);
        \draw[Arrow]    (a3) -- (a3 |- c.north);
        \draw[Arrow]    (a4) -- (a4 |- c.north);
        \draw[Arrow]    (a5) -- (a5 |- c.north);
        \draw[Arrow]    (b)  -- (b  |- c.north);
        \draw[Arrow]    (l1) -- (c.west  |- l1);
        \draw[Arrow]    (l2) -- (c.west  |- l2);
        \draw[Arrow]    (c.east  |- r) -- (r);
    \end{tikzpicture}
\caption{Disturbance factors on the measurement system.}\label{fig:DisturbanceFactors}
\end{figure}
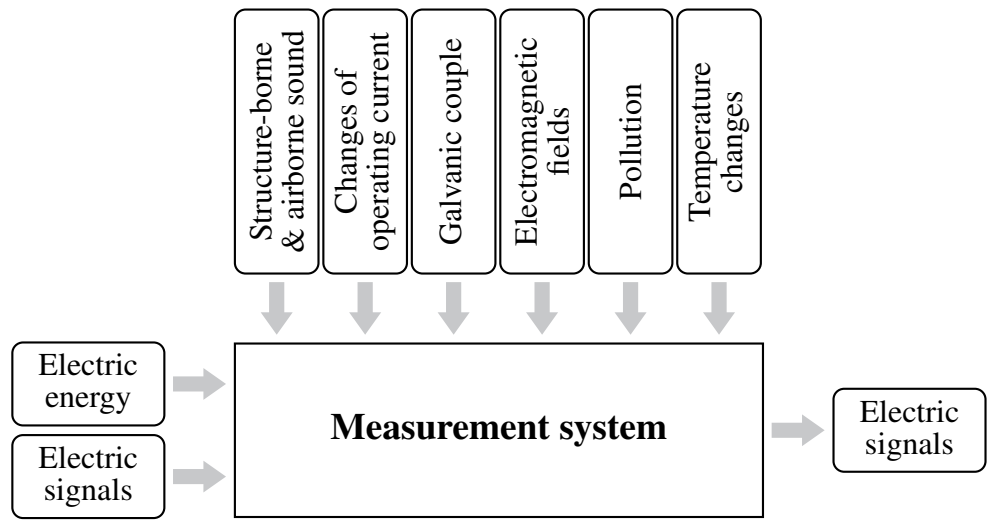

\subsubsection{Disturbance evaluation}
\begin{figure}[h!]
\begin{center}
\includegraphics[width=.8\linewidth]{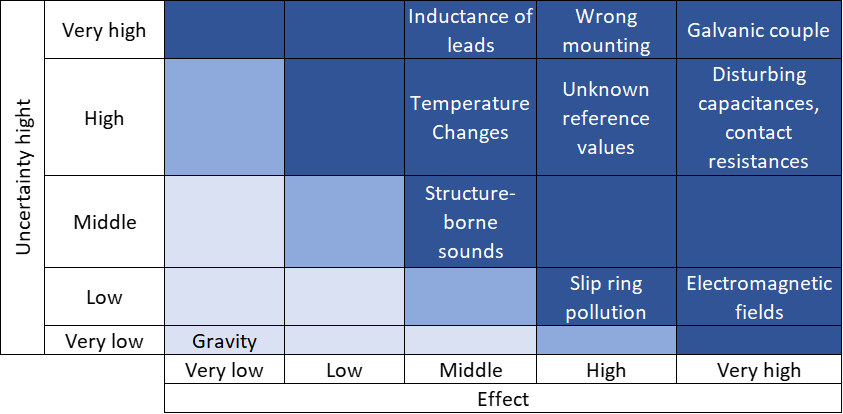}
\end{center}
\caption{Disturbance factor assessment.}\label{fig:DisturbanceFactorsAssessment}
\end{figure}

Structure-borne and airborne noise have a medium uncertainty level and a medium effect, since they reach the measurement point over longer distances or components with damping material properties lie in between. A direct influence on the measurements cannot be excluded. The impact is classified as medium, since vibrations due to structure-borne and airborne noise are part of the operating conditions of bearings. 
Since the slip ring is directly in the signal flow, the effects due to contamination on the measurement signal are high. However, the uncertainty level is low because contaminants are removed by rotation as a result of operation.
Electromagnetic fields are able to induce currents \citep{Schwab.2007}, so their effects are classified as very high. However, since there are distances through air between the sources and the measurement system, the uncertainty level is considered low. Changes in temperature can result in changes in resistance, which directly influence the electrical behavior of the measuring section \citep{Muhl.2020}. Therefore, they have a medium effect on the measurement quality, whereas the uncertainty level is to be classified as high. The reason for this is the non-enclosed environment of the test stand and the measuring system, whereby changes in the environmental influences have a direct effect on the systems. Unknown or not fully known reference variables are to be classified as high in amount and impact. Since they are directly included in the evaluation, all uncertainties associated with them are therefore also included in the result. 
Interference capacitance and contact resistances have a direct effect on the electrical path and can thus cause undefined current flows that directly influence the electrical behavior. Therefore, their effects on the measurement are very high. Due to the test chamber design, their uncertainty level is considered high. 
Line inductance and the associated line resistances occur in almost all test leads, and their uncertainty level must therefore be classified as very high. Their effects are not negligible because they conduct the electrical signals to the evaluation equipment.
Incorrect mounting of bearings has to be classified high in its effect, because it may cause changed operating conditions, which do not correspond to the rule. In case of insufficient mounting skills and knowledge, their uncertainty level is very high. As the previous evaluations make clear, uncertainties acting directly in or on the signal flow are to be considered very high in their effects. Therefore, undefined current flows, grounding problems and also damaged or insufficient insulation are extremely critical in their effect. They allow for changes in the electrical operating behavior, making the models on which the measurement is based only partially or no longer reflect the behavior. Since undefined current flows as a result of several factors and a grounding problem are given by the current technical status of the test chamber, its uncertainty level is very high. This also applies in the case of insulation, if damage is present. Figure~\ref{fig:DisturbanceFactorsAssessment} summarizes the results of the evaluation. 

\subsubsection{Disturbance handling}
In the following, only disturbance variables that have been classified as critical or no longer acceptable according to the Zurich hazard analysis method are considered with regard to their controllability. In order to find approaches for dealing with the disturbance variables, Mathias' catalog of principles is used at this point \citep{Mathias.2016}. Priority is given to the elimination of the disturbance variable or its influence, since an influence on the measurement can thus be excluded and consideration in the evaluation is no longer necessary.

\subparagraph*{Material-bound waves}
Structure-borne sound waves due to adjacent test equipment can be reduced or suppressed by decoupling the rolling bearing test rig from the ground \citep{Mathias.2016}. This is already done. The extent to which the effects on measurement were prevented cannot be assessed at this point.

\subparagraph*{Electricity and magnetism}
To reduce the influence of electromagnetic fields, it is recommended to shield the electrical conductors of the measuring system or to use shielded cables \citep{Mathias.2016}. Currently, \ac{bnc} cables are mainly used as measuring lines. They are a form of coaxial cable and are suitable for the low-loss transmission of high-frequency signals \citep{Rossmann.2018}. However, they exhibit parasitic effects which cannot be neglected \citep{Schwab.2007}. Therefore, these must be taken into account in the evaluation of the impedance signals. Another problem with coaxial cables is the effect they have on ground loops \citep{Schwab.2007}. Therefore, appropriate countermeasures must be taken. In order to reduce the influence of ground loops, a separate housing grounding of the test chamber is provided for the endurance tests. This is not grounded via a mains connection, but connected to a grounding bar subsequently installed in the test field. In this way, interference currents from the enclosure are to be discharged directly via the building ground and only have a minor effect on the carrier signal. The effect of this consideration will be checked in the course of this work in later tests.

\subparagraph*{Thermodynamics}
All three systems considered are subject to the same climatic conditions. Therefore, the influences on the impedance measurement caused by temperature changes are to be taken into account and compensated by the measurement and evaluation procedure. Measuring bridges generally have the advantage that they can compensate for disturbing influences such as temperature changes and also have a high measuring sensitivity \citep{Muhl.2020}. In addition, the compensation of the resistances of the \ac{bnc} lines is provided in the evaluation algorithm.

\subsection{Simulation Tests}
First, the frequency range in which capacitance changes can be measured reliably is investigated. Figure~\ref{fig:freqTest} shows simulation results of the Matlab/Simulink model described in Section~\ref{sec:simulationModel}. The capacitance was sinusoidally varied with a mean value of \SI{500}{\pico\farad}, an amplitude of \SI{400}{\pico\farad} and a varied frequency of \qtyrange{10}{500}{\hertz} as depicted on the $x$-axis. The $y$-axis shows the results of the peak simulated measurements as triangles and the true peaks of \SI{100}{\pico\farad} and \SI{900}{\pico\farad} as lines. It can be shown that the relative deviation $w_C = (C_\mathrm{real}-C_\mathrm{meas})/C_\mathrm{real}$ of the real values at a tenth of the balance capacitance is linearly dependent on the generator frequency $f_\mathrm{gen}$. It can be shown that this error can be approximated for a given capacitance change frequency $f_C$ as

\begin{equation}
    w_C = 1000 \cdot \left(\frac{f_C}{f_\mathrm{gen}}\right)^2 .
\end{equation}
Thus, a limit frequency can be derived for a maximum error of \SI{5}{\percent} at $f_C=f_\mathrm{gen}/141$ or for a maximum error of \SI{0.1}{\percent} at $f_C=f_\mathrm{gen}/1000$.

\begin{figure}[h!]
\begin{center}
    \begin{tikzpicture}
    
    \definecolor{clr1}{rgb}{0,0.3,0.45}
    \definecolor{clr2}{rgb}{.6,.75,0}
    \definecolor{clr3}{rgb}{.93,.4,0}
    
    \begin{semilogxaxis}[
        GenPlotStyle,
        width=0.5\linewidth,
        height=0.3\linewidth,
        xmin=10, xmax=10000,
        ymin=0, ymax=1000,
        xlabel={Frequency of capacitance change in \unit{\hertz}},
        ylabel={Measured capacitance peak in \unit{\pF}},
        legend style={at={(1.05,0.5)},anchor=west,legend cell align=left},
        no markers,
        legend entries={Minimum measured at $f_\mathrm{gen}=\SI{5}{\kilo\hertz}$,
                        Maximum measured at $f_\mathrm{gen}=\SI{5}{\kilo\hertz}$,
                        Minimum measured at $f_\mathrm{gen}=\SI{25}{\kilo\hertz}$,
                        Maximum measured at $f_\mathrm{gen}=\SI{25}{\kilo\hertz}$,
                        Minimum measured at $f_\mathrm{gen}=\SI{100}{\kilo\hertz}$,
                        Maximum measured at $f_\mathrm{gen}=\SI{100}{\kilo\hertz}$,
                        Real minimum,
                        Real maximum}];

    \addplot[clr1,only marks,mark options={mark=triangle,scale=2}]
         table[x=Frequency_of_capacitance_change_in_Hz, y=Minimum_measured_capacitance_in_pF_at_5kHz,col sep=comma] {figs/freqTest_V2.csv};

    \addplot[clr1,only marks,mark options={mark=triangle,rotate=180,scale=2}]
         table[x=Frequency_of_capacitance_change_in_Hz, y=Maximum_measured_capacitance_in_pF_at_5kHz,col sep=comma] {figs/freqTest_V2.csv};
         
    \addplot[clr2,only marks,mark options={mark=triangle,scale=2}]
         table[x=Frequency_of_capacitance_change_in_Hz, y=Minimum_measured_capacitance_in_pF_at_25kHz,col sep=comma] {figs/freqTest_V2.csv};

    \addplot[clr2,only marks,mark options={mark=triangle,rotate=180,scale=2}]
         table[x=Frequency_of_capacitance_change_in_Hz, y=Maximum_measured_capacitance_in_pF_at_25kHz,col sep=comma] {figs/freqTest_V2.csv};

    \addplot[clr3,only marks,mark options={mark=triangle,scale=2}]
         table[x=Frequency_of_capacitance_change_in_Hz, y=Minimum_measured_capacitance_in_pF_at_100kHz,col sep=comma] {figs/freqTest_V2.csv};

    \addplot[clr3,only marks,mark options={mark=triangle,rotate=180,scale=2}]
         table[x=Frequency_of_capacitance_change_in_Hz, y=Maximum_measured_capacitance_in_pF_at_100kHz,col sep=comma] {figs/freqTest_V2.csv};

    \addplot[black] coordinates {(10,100) (5179,100)};
    \addplot[black] coordinates {(10,900) (5179,900)};

    \end{semilogxaxis}
    \end{tikzpicture}
\end{center}
\caption{Simulation results of a sinusoidal changing capacitance $C_\mathrm{DUT}=\qtyrange{100}{900}{\pico\farad}$ at a rate of \qtyrange{10}{5000}{\hertz} measured with a generator frequency of $f_\mathrm{gen}=\qtylist{5;25;100}{\kilo\hertz}$. Displayed are the minimal and maximal values of the sinusoidal capacitance retrieved from the simulated measurement with the sampling frequency $f_\mathrm{s}=36 \cdot f_\mathrm{gen}$.}\label{fig:freqTest}
\end{figure}

\subsection{Single Contact Test}
\label{sec:single_contact_test}

Measuring a single steel ball deep groove ball bearing (6205) with the alternating voltage measurement bridge under load results in two voltage signals, which are depicted in Figure~\ref{fig:raw_single_contact}. The voltage signal $v_\mathrm{gen}$ shows the chosen sinusoidal curve with an amplitude of \SI{6}{\volt} and a frequency of \SI{20}{\kilo\hertz}. The measured voltage signal $\underline{v}_\mathrm{m}$ shows the same frequency and no phase shift, with a lower amplitude of approximately \SI{1.15}{\volt}. The reduced amplitude results from the electrical circuit of the used measurement bridge (Figure~\ref{fig:ac-bridge_circuit}) combined with the altered electrical behavior of the single steel ball bearing. 

\begin{subfigure}
    \setcounter{subfigure}{0}
    \centering
    \begin{minipage}[t]{0.45\textwidth}
        \begin{tikzpicture}
        
        \definecolor{clr1}{rgb}{0,0.3,0.45}
        \definecolor{clr2}{rgb}{.6,.75,0}
        
        \begin{axis}[
            GenPlotStyle,
            width=\linewidth,
            ylabel={Raw voltage signal in \unit{\volt}},
            xlabel={Time in \unit{\milli\second}},
            xmin=0,
            ]
        
        \addplot[clr1]
            table[x=xAxis, y=ugen, col sep=comma] {figs/ResultsRawMeas.csv};
        \addlegendentry{$v_\mathrm{gen}$}
        
        \addplot[clr2]
            table[x=xAxis, y=umess,col sep=comma] {figs/ResultsRawMeas.csv};
        \addlegendentry{$v_\mathrm{m}$}
        
        \end{axis}
    \end{tikzpicture}
\caption{Generator and measured voltage signal.} 
\label{fig:raw_single_contact}
    \end{minipage}  
    \hfill
\setcounter{subfigure}{1}
    \begin{minipage}[t]{0.45\textwidth}
       \begin{tikzpicture}
        
        \definecolor{clr1}{rgb}{0,0.3,0.45}
        
        \begin{axis}[
            GenPlotStyle,
            width=\linewidth,
            ylabel={Capacitance in pF},
            xlabel={Time in \unit{\second}},
            xmin=0,
            ]
        
        \addplot[clr1]
            table[x=xAxis, y=C_L, col sep=comma] {figs/ResultsCMeas.csv};
        
        \end{axis}
    \end{tikzpicture}
    \caption{Calculated capacitance signal from the raw voltage signals.}
    \label{fig:CL_time}
    \end{minipage}

\setcounter{subfigure}{-1}
    \caption{Raw (A) and calculated (B) results of a single steel ball bearing at $T_\mathrm{Oil}=\SI{30}{\celsius}$, $n=\SI{1000}{\per\minute}$, $f_\mathrm{gen}=\SI{20}{\kilo\hertz}$, and $F_\mathrm{r}=\SI{2525}{\newton}$.}
    \label{fig:circuitDiagrams}
\end{subfigure}

        
        
        
        
        

Using the proposed methods of Section~\ref{sec:measurment_bridge_function} the capacitance signal over time can be calculated from the raw $v_\mathrm{gen}$ and $\underline{v}_\mathrm{m}$ signals (Figure~\ref{fig:CL_time}). Each cage rotation can be easily separated by a significant peak in capacitance. As the steel ball reaches the load zone, the gap between the steel surfaces of the ball and raceway gets smaller and thus the capacitance increases. Outside the load zone, the geometrically restricted gap between the ball and raceway induces a minimum capacitance of \SI{18}{\pico\farad}. The noise and variation observed when comparing each rotation stem from system distortions, as discussed in Section~\ref{sec:umea_results}. 

Separating each rotation of the ball bearing, an averaged capacitance signal can be extracted for different load conditions. In Figure~\ref{fig:MultiLoad} multiple averaged capacitance measurements are depicted with varying radial force $F_\mathrm{r}$ and constant oil temperature $T_\mathrm{Oil}$, rotation speed $n$ and measurement frequency $f_\mathrm{gen}$. Increasing radial forces result in higher capacitances inside the load zone, as already discussed. However, capacitances outside the load zone are decreasing with rising radial forces. This effect is caused by the deflection of the balls under load. Higher loads result in a more deflected ball inside the load zone, thus more space on the other side of the ball bearing is present. This larger gap induces a lower capacitance compared to ball bearings under less radial force.

        
        
        
        


\begin{figure*}[htb]
\centering

    \usepgfplotslibrary{fillbetween}
    \begin{tikzpicture}
    
    \definecolor{clr1}{rgb}{0,0.3,0.45}
    \definecolor{clr2}{rgb}{.6,.75,0}
    \definecolor{clr3}{rgb}{.99,.79,0}
    \definecolor{clr4}{rgb}{0,.51,.80}
    \definecolor{clr5}{rgb}{.92,.39,0}
    \definecolor{clr6}{rgb}{.9,0,.1}
    
    \begin{axis}[
        GenPlotStyle,
        width = 210pt,
        height = 200pt,
        xlabel={Time in \unit{\second}},
        ylabel={Averaged capacitance in \unit{\pF}}, 
        legend style={at={([xshift=3pt,yshift=0pt]1,0.5)},anchor=west,legend cell align=left},
        legend image code/.code={
            \draw[mark repeat=2,mark phase=2]
            plot coordinates {
            (0cm,0cm)
            (0.15cm,0cm)        
            (0.3cm,0cm)         
            };%
            },
        legend entries={std,
                        \SI{994}{\N},
                       \SI{1590}{\N},
                       \SI{2525}{\N},
                       \SI{3975}{\N},
                       \SI{6360}{\N}
               },
        reverse legend]
    ]
    \addlegendimage{only marks, mark=square*,color=gray, opacity=0.6} 
        
    \addplot[clr1]
         table[x=xAxis, y=mean1,col sep=comma] {figs/Results5C.csv};
    
    \addplot[clr1, opacity=0,name path=u1, forget plot]
         table[x=xAxis, y=upper_std1,col sep=comma] {figs/Results5C.csv};
         
    \addplot[clr1, opacity=0,name path=l1, forget plot]
         table[x=xAxis, y=lower_std1,col sep=comma] {figs/Results5C.csv};
         
    \addplot[clr1, opacity=0.2, forget plot] fill between[of=u1 and l1];
    
    \addplot[clr2]
         table[x=xAxis, y=mean2,col sep=comma] {figs/Results5C.csv};
    
    \addplot[clr2, opacity=0,name path=u2, forget plot]
         table[x=xAxis, y=upper_std2,col sep=comma] {figs/Results5C.csv};
         
    \addplot[clr2, opacity=0,name path=l2, forget plot]
         table[x=xAxis, y=lower_std2,col sep=comma] {figs/Results5C.csv};
         
    \addplot[clr2, opacity=0.2, forget plot] fill between[of=u2 and l2];
    
    \addplot[clr3]
         table[x=xAxis, y=mean3,col sep=comma] {figs/Results5C.csv};
    
    \addplot[clr3, opacity=0,name path=u3, forget plot]
         table[x=xAxis, y=upper_std3,col sep=comma] {figs/Results5C.csv};
         
    \addplot[clr3, opacity=0,name path=l3, forget plot]
         table[x=xAxis, y=lower_std3,col sep=comma] {figs/Results5C.csv};
         
    \addplot[clr3, opacity=0.3, forget plot] fill between[of=u3 and l3];

    \addplot[clr4]
         table[x=xAxis, y=mean4,col sep=comma] {figs/Results5C.csv};
    
    \addplot[clr4, opacity=0,name path=u4, forget plot]
         table[x=xAxis, y=upper_std4,col sep=comma] {figs/Results5C.csv};
         
    \addplot[clr4, opacity=0,name path=l4, forget plot]
         table[x=xAxis, y=lower_std4,col sep=comma] {figs/Results5C.csv};
         
    \addplot[clr4, opacity=0.5, forget plot] fill between[of=u4 and l4];

    \addplot[clr6]
         table[x=xAxis, y=mean5,col sep=comma] {figs/Results5C.csv};
    
    \addplot[clr6, opacity=0,name path=u5, forget plot]
         table[x=xAxis, y=upper_std5,col sep=comma] {figs/Results5C.csv};
         
    \addplot[clr6, opacity=0,name path=l5, forget plot]
         table[x=xAxis, y=lower_std5,col sep=comma] {figs/Results5C.csv};
         
    \addplot[clr6, opacity=0.2, forget plot] fill between[of=u5 and l5];
    
    \end{axis}
    \end{tikzpicture}

\caption{Averaged capacitance measurements of single steel ball bearing at $T_\mathrm{Oil}=\SI{30}{\celsius}$, $n=\SI{1000}{\per\minute}$, $f_\mathrm{gen}=\SI{20}{\kilo\hertz}$, and multiple forces.}
\label{fig:MultiLoad}
\end{figure*}

\subsection{Fatigue Tests}
Investigations of the data generated by using the alternating current measurement bridge show that three different phases can be identified in bearings' operating life in the frequency domain. 
According to Martin et al., a run-in phase, a normal operation phase and a failure phase can be detected, which ensures the usage of the new measurement approach \cite{Martin.2022}. In time domain, the real part of the impedance data of Martin et al. is negative \cite{Martin.2022}. In contrast, the real part of the signals generated with this measurement approach is not negative. That makes sense, because a negative real part would implicate an energy gain in the system, which is physically implausible. In general, the measured signals have a high repeatability in their absolute value under the same operating conditions. Comparing different bearing types and different load angles, it can be said that the results are independently of the bearing type. In case of additional axial load, the signal changes indicating bearing's operational life end occur later compared to pure radial loads. But the significance of the signal changes still enables a damage detection using this approach. Therefore, the alternating measurement bridge can be used as an observation approach \cite{BeckerDombrowsky.2023}. 

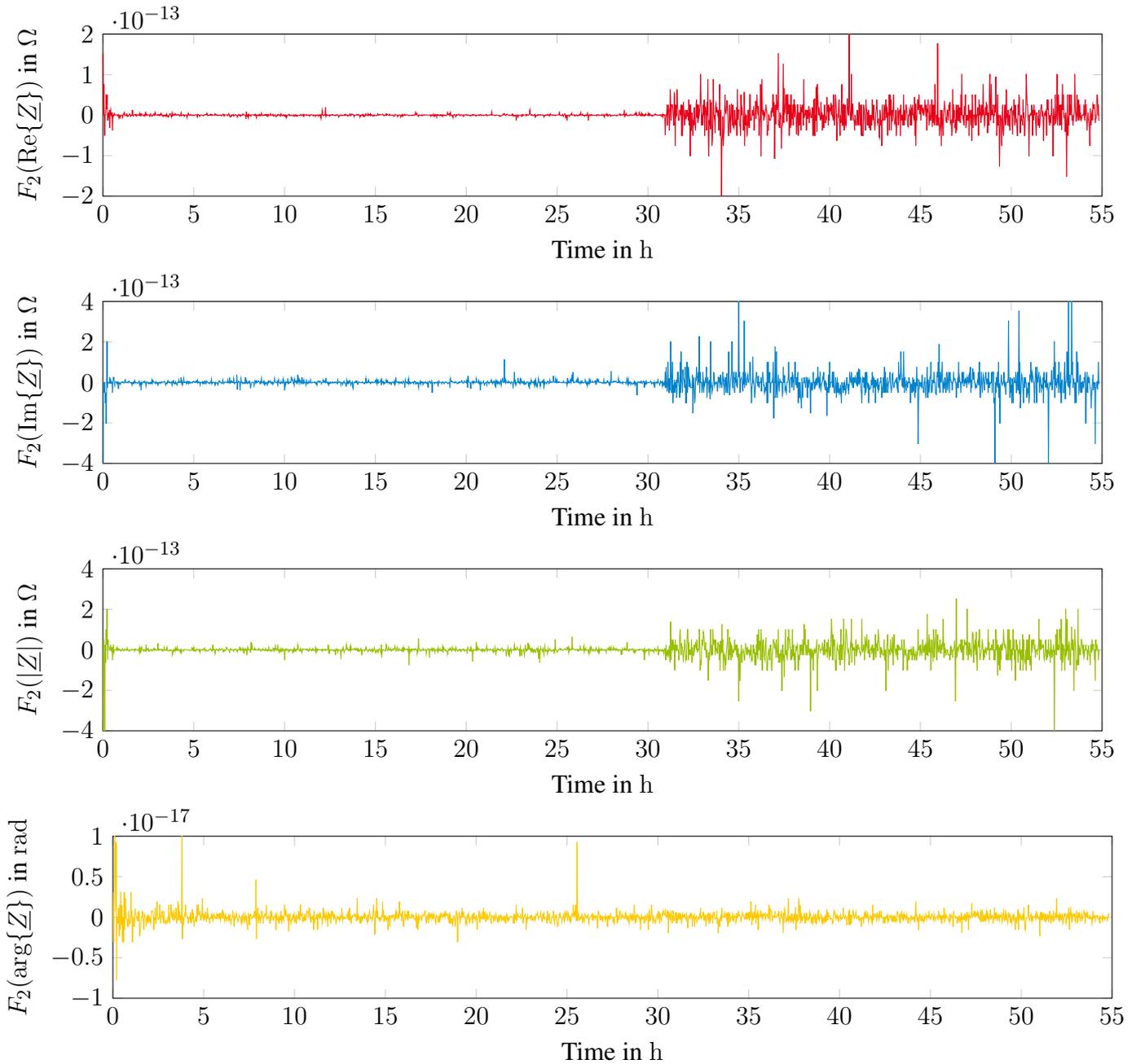
\begin{figure}[h!]
     \centering
     \begin{tabular}{c}
         \begin{tikzpicture}
             \definecolor{clr6}{rgb}{.9,0,.1}
             \begin{axis}[
                 xlabel={Time in \unit{\hour}},
                 ylabel={$F_2(\operatorname{Re}\{\underline{Z}\})$ in \unit{\ohm}},
                 xmin=0, xmax=55,
                 ymin=-2e-13, ymax=2e-13,
                 width=\textwidth,
                 height = 120pt
             ] 
             \addplot[clr6]
              table[x=Zeit, y=Re_F_2_imag,col sep=comma] {figs/M_01-La_A_modifiziert.csv};
             \end{axis}
         \end{tikzpicture}
         \\
         \begin{tikzpicture}
             \definecolor{clr4}{rgb}{0,.51,.80}
             \begin{axis}[
                 xlabel={Time in \unit{\hour}},
                 ylabel={$F_2(\operatorname{Im}\{\underline{Z}\})$ in \unit{\ohm}},
                 xmin=0, xmax=55,
                 ymin=-0.4e-12, ymax=0.4e-12,
                 width=\textwidth,
                 height = 120pt
             ] 
             \addplot[clr4]
              table[x=Zeit, y=Im_F_2_imag,col sep=comma] {figs/M_01-La_A_modifiziert.csv};
             \end{axis}
         \end{tikzpicture}
         \\
         \begin{tikzpicture}
             \definecolor{clr2}{rgb}{.6,.75,0}
             \begin{axis}[
                 xlabel={Time in \unit{\hour}},
                 ylabel={$F_2(|\underline{Z}|)$ in \unit{\ohm}},
                 xmin=0, xmax=55,
                 ymin=-0.4e-12, ymax=0.4e-12,
                 width=\textwidth,
                 height = 120pt
             ] 
             \addplot[clr2]
              table[x=Zeit, y=abs_F_2_imag,col sep=comma] {figs/M_01-La_A_modifiziert.csv};
             \end{axis}
         \end{tikzpicture}
         \\
         \begin{tikzpicture}
             \definecolor{clr3}{rgb}{.99,.79,0}
             \begin{axis}[
                 xlabel={Time in \unit{\hour}},
                 ylabel={$F_2(\operatorname{arg}\{\underline{Z}\})$ in \unit{\radian}},
                 xmin=0, xmax=55,
                 ymin=-1e-17, ymax=1e-17,
                 width=\textwidth,
                 height = 120pt
             ] 
             \addplot[clr3]
              table[x=Zeit, y=phase_F_2_imag,col sep=comma] {figs/M_01-La_A_modifiziert.csv};
             \end{axis}
         \end{tikzpicture}
     \end{tabular}

\caption{Frequency feature $F_2$ - central frequency - for real part ($\operatorname{Re}\{\underline{Z}\}$), imaginary part ($\operatorname{Im}\{\underline{Z}\}$), absolute value ($|\underline{Z}|$) and phase angle ($\operatorname{arg}\{\underline{Z}\}$) over operational time of bearing A, bearing type 6205, radial load $F_\mathrm{r}=\SI{9375}{\newton}$, load ratio  $C/P = \num{1.6}$, load angle $\beta = \SI{0}{\degree}$, rotational speed $n=\SI{5000}{rpm}$, oil flow of \SI{10}{\litre \per\minute}.}
\label{fig:M_01}
\end{figure}

Investigating the signals in time and frequency domain increases the information density for condition monitoring. It is essential to detect the damage phase and to differentiate between run-in phase and damage phase. As described by Martin et al. and Becker-Dombrowsky at al., the run-in phase is dominant in the pure impedance signal over the operational time \citep{Martin.2022, BeckerDombrowsky.2023}. Therefore, different approaches are used to optimize the signal. To do so, the impedance signal is transformed into frequency domain and different features have been calculated and plotted over the operational time to identify significant signal changes. Figure~\ref{fig:M_01} shows exemplary the behavior of the frequency feature $F_2$, the central frequency, according to Table~2 of Becker-Dombrowsky et al. for the real part ($\operatorname{Re}\{\underline{Z}\}$), imaginary part ($\operatorname{Im}\{\underline{Z}\}$), absolute value ($|\underline{Z}|$) and phase angle ($\operatorname{arg}\{\underline{Z}\}$) over the test time \citep{BeckerDombrowsky.2023}. 

The feature signals of real part, imaginary part and absolute value behave different for run-in phase, normal condition phase and damage phase. Due to a higher number of metallic contacts in the run-in phase, the phase angle tends temporary to zero and back again to $\SI{-90}{\degree}$. Therefore, the central frequency changes significantly at the first hours, which can be seen in the yellow graph in Figure~\ref{fig:M_01}.This allows the differentiation between these three phases and enables the phase classification using machine learning approaches. In addition, the feature changes significantly hours before the test was stopped by the test-rig internal vibration observation, which indicates the possibility of a damage early detection before vibration observation detects damages. Therefore, the alternating current measurement bridge approach can be seen as an instrument for damage early detection for rolling element bearings.

\section{Conclusions}
In this work, a  visualization technique for rolling bearings, the impedance measurement, was improved. A suitable measurement principle was identified and developed to measure the impedance of rolling element bearings at a several \unit{\kilo\hertz} sampling rate and high disturbance resilience. First, different impedance measurement principles were introduced and an \ac{umea} was carried out for the most promising approach, the unbalanced AC Wheatstone bridge. A simulation model was developed to study disturbance on the measurement and to investigate the edge frequency up to which a frequency domain analysis of the measured impedance is meaningful. To estimate the latter, an equation was derived. Then, the measurement bridge was physically build and tested in two practical applications. Firstly, measuring the impedance of single steel ball bearings to measure the contact capacitance of a singe ball in a radial deep groove ball bearing. Secondly, the impedance of a normal radial deep groove ball bearing was measured over its lifetime in an accelerated fatigue test. In both tests, the AC Wheatstone bridge delivered low-noise and high-frequency impedance signals and therefore proofed suitable for its application in measuring rolling bearing's impedance. 

\section*{List of abbreviations}

\begin{acronym}[UMEA ] 
    \acro{ac}[AC]{alternating current}
    \acro{bnc}[BNC]{bayonet nut connector}
    \acro{dut}[DUT]{device under test}
    \acro{ehl}[EHL]{elasto hydrodynamic lubrication}
    \acro{umea}[UMEA]{uncertainty mode and effects analysis}
\end{acronym}

\bibliographystyle{Frontiers-Harvard} 

\bibliography{refs_Pu,refs_Bk}

\end{document}